\documentclass[showpacs,preprintnumbers,amsmath]{revtex4}
\usepackage{amssymb}

\usepackage{graphicx}
\usepackage{dcolumn}
\usepackage{bm}

\begin{document}

\title{Unified analytical treatments to qubit-oscillator systems}
\date{\today}
\author{Shu He$^{1}$, Yu-Yu Zhang$^{2}$, Qing-Hu Chen$^{3,4,*}$, Xue-Zao Ren$^{1}$,
Tao Liu$^{1}$, and Ke-Lin Wang$^{5}$}

\address{
$ ^{1}$ School of Science,  Southwest University of  Science and
Technology, Mianyang 621010, P.  R.  China\\
$^{2}$Center for Modern Physics, Chongqing University, Chongqing
400044, P. R.  China\\
$^{3}$ Department of Physics, Zhejiang University, Hangzhou 310027,
P. R. China \\
$^{4}$ Center for Statistical and Theoretical Condensed Matter
Physics, Zhejiang Normal University, Jinhua 321004, P. R. China  \\
$^{5}$ Department of Modern Physics, University of  Science and
Technology of China,  Hefei 230026, P.  R.  China
 }
\date{\today}

\begin{abstract}
An effective scheme within two displaced bosonic operators with
equal positive and negative displacements is extended to study
qubit-oscillator systems analytically in an unified way. Many
previous analytical treatments, such as generalized rotating-wave
approximation (GRWA) [Phys. Rev. Lett. \textbf{\ 99}, 173601 (2007)]
and an expansion in the qubit tunneling matrix element in the deep
strong coupling regime [Phys. Rev. Lett. \textbf{105}, 263603
(2010)] can be recovered straightforwardly within the present
scheme. Moreover, further improving GRWA and extension to the
finite-bias case are implemented easily.  The analytical expressions
are then derived explicitly and uniquely, which work well  in a wide
range of the coupling strengthes, detunings, and static bias
including the recent experimentally accessible parameters.
\end{abstract}

\pacs{42.50.Lc, 42.50.Pq, 32.30.-r, 03.65.Fd}

\maketitle

\section{introduction}

Matter-matter interaction is fundamental and ubiquitous in modern
physics ranging from quantum optics, quantum information science  to
condensed matter physics.  The simplest paradigm is a two-level atom
(qubit) coupled to the electromagnetic mode of a cavity
(oscillator). In the strong coupling regime where the coupling
strength $ g/\omega $ ($\omega $ is the cavity frequency) between
the atom and the cavity mode exceeds the loss rates, the atom and
the cavity can repeatedly exchange excitations before coherence is
lost. The Rabi oscillations can be observed in this strong coupling
atom-cavity system, which is usually called as cavity quantum
electrodynamics (QED) \cite{CQED}. Typically, the coupling strength
in cavity QED reaches $g/\omega \thicksim 10^{-6}$. It can be
described by the well-known Jaynes-Cummings (JC) model~ \cite{JC}
without the rotating-wave approximation (RWA).

Recently, for superconducting qubits, a one-dimensional (1D) transmission
line resonator~\cite{Wallraff} or a LC circuit~\cite
{Chiorescu,Wang,Deppe,Fink} can play a role of the cavity, which is known
today as circuit QED. More recently, LC resonator inductively coupled to a
superconducting qubit~\cite{Niemczyk,exp,Mooij} has been realized
experimentally. The qubit-resonator coupling has been strengthened from $%
g/\omega \thicksim 10^{-3}$ in the earlier
realization~\cite{Wallraff}, a few percentage
later~\cite{Fink,Deppe} , to most recent ten percentages~\cite
{Niemczyk,exp,Mooij}. Due to the ultra-strong coupling strength
$g/\omega \thicksim 0.1$, evidence for the breakdown of the RWA has
been provided~\cite{Niemczyk}. Recently, some works have been
devoted to this qubit-oscillator system in the ultra-strong coupling
regime \cite{Werlang,Hanggi,Nori,Hausinger,Qinghu}.

Actually, the JC model  without the RWA in a wide coupling regime
has been studied extensively for more than 40 years. By
polaronic-like transformations or displaced operators, various
analytical and numerical approaches have been developed in recent
years, an  incomplete list is given by Refs \cite {Irish,chenqh,liu,chen10,Zheng,Casanova,QingHu1,Braak}. 
Very accurate or exact solutions have been obtained.
Various forms of RWA energies have been developed
\cite{Albert,Jonasson} and extensions to the N-level case have been
performed recently\cite{Albert1}. Most works have been mainly devoted to the zero static bias. 
The  analytical expression 
with high accuracy  for the qubit-Oscillator systems with both
zero and finite static bias should be of practical interest.  

In this paper, by using two displaced bosonic operators with equal
positive and negative displacements, we can recover many previous
analytical treatments for both zero and finite static bias within
the same scheme.
What is more, we can extend analytically the previous generalized
RWA (GRWA) to the finite static bias. Beyond the GRWA for the zero
bias is also performed. The expression is uniquely given and the
results are very close to exact ones for wide range of the model
parameters which cover the present-day experimentally accessible
parameters.

\section{Model and exact solution}

The Hamiltonian for a superconducting qubit coupled to a harmonic oscillator
in circuit QED consists of three parts\cite{Nori,Hausinger}. The first one
is the interaction between the qubit and the LC resonator, which is
described by
\begin{equation}
H_{int}=\hbar g(a^{\dagger }+a)\sigma _z,
\end{equation}
where $a^{\dagger }$, $a$ are the photon creation and annihilation operators
in the basis of Fock states of the LC resonator, $g$ is the qubit-cavity
coupling constant. The RWA has not been employed here. The effective
Hamiltonian for the qubit can be written as the standard one for a two-level
system
\begin{equation}
H=-\frac 12\left( \varepsilon \sigma _z+\Delta \sigma _x\right),
\end{equation}
where $\varepsilon $ and $\Delta $ are qubit static bias and tunneling
matrix element. In the recent circuit QED \cite{Niemczyk,exp} operating in
the ultra-strong coupling regime, they describe the transition frequency of
the flux qubit and the tunneling coupling between the two persistent current
states. $\varepsilon =I_p(\Phi -\Phi _0/2)$ with $I_p$ the persistent
current in the qubit loop, $\Phi $ an externally applied magnetic flux, and $%
\Phi _0$ the flux quantum. In contrast to atomic cavity QED systems, $%
\varepsilon$ is easily tunable in circuit QED systems using superconducting
qubit. In the above two equations, the Pauli matrix notations $\sigma
_k(k=x,y,z)$ $\ $ are used in the basis of the two persistent current
states. The third one is LC resonator $\omega a^{\dagger }a$ with single
mode frequency $\omega$. Then the Hamiltonian for the whole system reads $%
\left( \hbar =\omega =1\right) $
\begin{equation}
H=-\frac 12\left( \varepsilon \sigma _z+\Delta \sigma _x\right) +
a^{\dagger }a+g(a^{\dagger }+a)\sigma _z  \label{Hamiltonian}.
\end{equation}

Motivated by the work in the Dicke model\cite{chenqh}, we have introduced
two displaced bosonic operators with equal positive and negative
displacements in this system\cite{Qinghu}
\begin{equation}
A=a+g,B=a-g  \label{shift},
\end{equation}
then the Hamiltonian can be written in the following matrix form
\begin{equation}
H=\left(
\begin{array}{cc}
A^{\dagger }A-g^2-\frac \varepsilon 2 & -\Delta /2 \\
-\Delta /2 & B^{\dagger }B-g^2+\frac \varepsilon 2
\end{array}
\right)   \label{Hamiltonian}.
\end{equation}
Note that the linear term for the original bosonic operator
$a^{\dagger }(a)$ is removed, and only the number operators $A^{+}A$
and $B^{+}B$ are left. Therefore the wavefunction can be expanded in
terms of these new operators as
\begin{equation}
\left| {}\right\rangle =\left(
\begin{array}{l}
\sum_{n=0}^{N_{tr}}c_n\left| n\right\rangle _A \\
\sum_{n=0}^{N_{tr}}(-1)^nd_n\left| n\right\rangle _B,
\end{array}
\right)   \label{wavefunction},
\end{equation}
where $N_{tr}$ is the truanted number. For $A$ operator, we have
\begin{eqnarray}
\left| n\right\rangle _A &=&\frac{\left( A^{\dagger }\right)
^n}{\sqrt{n!}} \left| 0\right\rangle _A=\frac{\left( a^{\dagger
}+g\right) ^n}{\sqrt{n!}} \left| 0\right\rangle _A, \\ \left|
0\right\rangle _A &=&e^{-\frac 12g^2-ga^{\dagger })}\left|
0\right\rangle _a,
\end{eqnarray}
$B$ operator has the same properties. Inserting Eqs. (6) and (7) into the
Schr$\stackrel{..}{o}$ dinger equation, we have
\begin{eqnarray}
\left( m-g^2-\frac \varepsilon 2\right) c_m-\sum_{n=0\ }D_{mn}d_n
&=&Ec_m,
\label{exact1} \\
\left( m-g^2+\frac \varepsilon 2\right) d_m-\sum_{n=0}D_{mn}c_n
&=&Ed_m \label{exact2},
\end{eqnarray}
where
\begin{eqnarray}
D_{mn} &=&\frac \Delta 2\left( -1\right) ^m{}_B\left\langle m\right| \left|
n\right\rangle _A, \\
_B\left\langle m\right| \left| n\right\rangle _A &=&(2g)^{n-m}\exp
(-2g^2)\sqrt{\frac{m!}{n!}}L_m^{n-m}\left( 4g^2\right),
\end{eqnarray}
for $n\ge m$, $L_m^{n-m}(x)$ is Laguerre polynomial,
$D_{mn}=D_{nm}$.

Based on Eqs. (\ref{exact1}) and (\ref{exact2}), we have given
numerically exact solutions to the qubit-oscillator system with any
finite static bias $ \varepsilon $\cite{Qinghu}. In this paper, we
alternatively present some analytical results in the framework of
the above formalism. One can see that some recent analytical results
by other authors are explicitly covered in the present framework.
Moreover, the present scheme is more convenient to perform further
analytical studies.

\section{Analytical treatments}

\subsection{Variational study for $\epsilon=0$}

To have a sense of two displaced bosonic operators Eq. (\ref{shift}), we
relax the displacement to be a variational parameter $\alpha $,
\begin{equation}
A=a+\alpha, B=a-\alpha,
\end{equation}
then study the unbiased Hamiltonian ($\varepsilon =0$) variationally.
Suppose that the trial state is the vacuum state in these displaced
operators as the following
\[
|\Phi _0\rangle =\left( \
\begin{array}{l}
\left| 0\right\rangle _A \\
\left| 0\right\rangle _B
\end{array}
\right).
\]
The energy expectation is derived as
\begin{equation}
E_0=-\frac \Delta 2e^{-2\alpha ^2}-2g\alpha +\alpha ^2  \label{var}.
\end{equation}
Minimizing the energy gives
\begin{equation}
\Delta \alpha e^{-2\alpha ^2}-g+\alpha =0.  \label{para}
\end{equation}

In the weak coupling limit, we can obtain the variational parameter
and the ground state (GS) energy respectively
\begin{equation}
\alpha =\frac g{1+\Delta },
\end{equation}
\begin{equation}
E_0=-\frac \Delta 2\exp \left[ -2\left( \frac g{1+\Delta }\right)
^2\right] -g^2\frac{1+2\Delta }{\left( 1+\Delta \right) ^2},
\end{equation}
which are exactly the same as Eqs. (7) and (8) obtained in Ref.
\cite {Yuanwei}.

In the strong-coupling limit, the first term in Eq. (\ref{var}), which is
originated from the qubit tunneling, is too small and can be neglected, then
we simply have
\begin{equation}
\alpha =g,
\end{equation}
and the GS energy
\begin{equation}
E^{SCL}=-g^2.
\end{equation}

For the arbitrary coupling, one can solve Eq. (\ref{para})
consistently and the reasonable GS energy will be derived, which is
not shown here.

\subsection{Perturbation theory based on the exact solution in the strong
coupling limit}

Note above that in the strong coupling limit, the variational parameter is
just exactly $\alpha =g$. It can be also readily obtained by neglecting the
qubit tunneling term $-\frac 12\Delta \sigma _x$ in Hamiltonian (\ref
{Hamiltonian}) with zero static-bias. In this case, based on the displaced
operators $A$ and $B$, the eigenstates are easily obtained as
\begin{equation}
\left| m\right\rangle ^{(0)}=\left(
\begin{array}{l}
\left| m\right\rangle _A \\
\pm (-1)^m\left| m\right\rangle _B\label{wf_scl}
\end{array}
\right),
\end{equation}
and the eigenvalues are $E_m^{\pm 0}=m-g^2$ for the $m$ state. Note that the
eigenstates are twofold degenerate.

Next, considering $H^{\prime }=-\frac 12\Delta \sigma _x$ as a perturbation,
within the second-order perturbation theory, we can readily derive the
eigenenergy with even (odd) parity for zero qubit static bias as
\begin{equation}
E_m^{\pm }=m-g^2\mp (-1)^mD_{mm}+\sum_{m\neq n}\frac{\left|
D_{mn}\right| ^2 }{m-n}  \label{DSC}.
\end{equation}
It is interesting to note that it is just the same as Eq. (5) in Ref. \cite
{Casanova} obtained by Casanova et al. in the deep strong coupling (DSC)
regime of the JC model.

\begin{figure}[tbp]
\includegraphics[width=16cm]{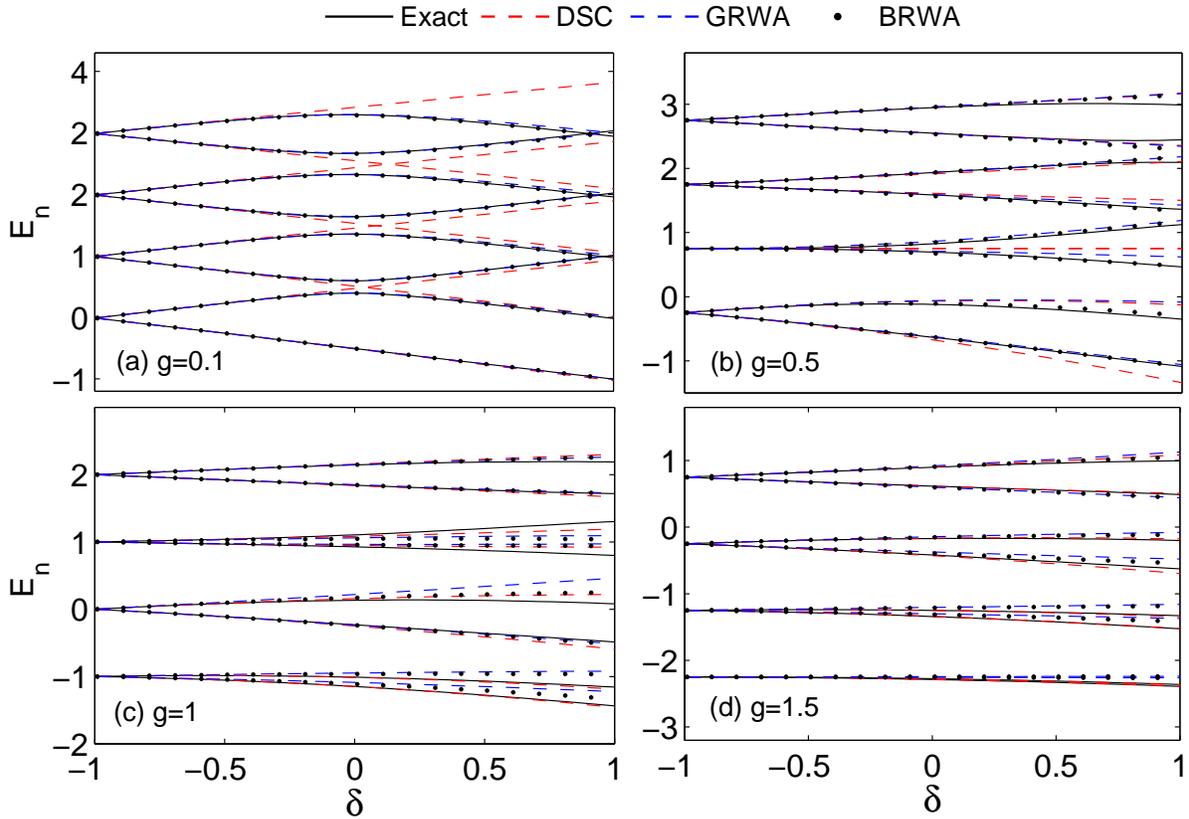}
\caption{ The energy levels for zero bias $\epsilon =0$ at (a)
$g=0.1$, (b) $ g=0.5$ , (c) $g=1.0$, and (d) $g=1.5$. The present
BRWA results (black filled circles) are compared to the exact (solid
lines), DSC(red dashed lines), and GRWA (blue  dashed lines) ones. }
\label{unbias}
\end{figure}

The energy levels by Eq. (\ref{DSC}) against the qubit-oscillator
detuning $ \delta =\Delta -1\;$ for $g=0.1,0.5,1.0$, and $1.5$,
ranging from weak to deep strong coupling, are displayed in Fig.
\ref{unbias}. The numerically exact results from Eq. (\ref{exact1})
and (\ref{exact2}) are also collected as a benchmark. It is found
that the DSC results are especially  suited to the DSC regime or
small detunings. Note that Casanova et al just focused on the
investigation in the DSC regime ($g/\omega =2$) or small detunings
($\Delta \le 0.5$). At weak coupling $g=0.1$, it is shown in Fig.
\ref{unbias} (a) that, even for the negative detuning $\delta$, the
DSC deviates from the exact ones. However, in the present
experimentally accessible systems, the maximum value for the
coupling strength is generally realized in the superconducting flux
qubit coupled to a circuit resonant \cite{Niemczyk}, which is only
around $ g=0.1$, to our knowledge. So it should be practically
interesting to find a good solution in this coupling regime.

For any value of the qubit bias $\varepsilon $, the Hamiltonian
(\ref {Hamiltonian}) with a vanishing tunneling element $\Delta =0$
can be diagonalized in terms of two eigenstates $\left| \uparrow
,m\right\rangle _A\ $ and $\left| \downarrow ,m\right\rangle _B$
with $\left| \uparrow \right\rangle $ ($\left| \downarrow
\right\rangle $) the eigenstate of $ \sigma _z$, the corresponding
eigenvalues are
\begin{equation}
E_{\uparrow /\downarrow ,m}^{(0)}=\pm \frac 12\varepsilon +m-g^2,
\end{equation}
For finite $\Delta $, the perturbative matrix elements becomes
\begin{equation}
-\frac 12\;_B\left\langle \downarrow ,m\right| \Delta \sigma
_x\left| \uparrow ,n\right\rangle _A=-\left( -1\right) ^mD_{mn}.
\end{equation}
Note that these two euqations are exactly the same as Eqs. (7), (8) in Ref.
\cite{Hausinger}. Then the full Hamiltonian can be diagonalized
perturbatively to second-order in $\Delta $ by using Van Vleck perturbation
(VVP) theory as outlined in Ref. \cite{Hausinger}. The eigenvalue is given
by
\begin{eqnarray}
E_m^{\pm } &=&m+\frac l2-g^2+\frac 12\sum_{k=0,k\neq m\pm l}^\infty \left(
\frac{D_{mk}^2}{\varepsilon +m-k}-\frac{D_{nk}^2}{\varepsilon +k-n}\right)
\nonumber \\
&&\mp \frac 12\sqrt{\left[ \varepsilon -l+\sum_{k=0,k\neq m\pm
l}^\infty \left( \frac{D_{mk}^2}{\varepsilon
+m-k}-\frac{D_{nk}^2}{\varepsilon +k-n}
\right) \right] ^2+4D_{mn}^2},  \nonumber \\
n &=&m+l(l\geq 0),m=0,1,2,...  \label{VVP}
\end{eqnarray}
which is the same as Eqs . (12) for VVP in Ref. \cite{Hausinger}. So
the VVP for finite bias can be also recovered easily in the present
scheme. It can be reduced to the zero-bias case Eq. (\ref{DSC}) by
set $\varepsilon =0\;(m=n)$.

It has been shown \cite{Hausinger} that VVP works very well in the
deep strong coupling or large static bias. It is consistent with the
fact that the unperturbative Hamiltonian includes the
qubit-oscillator interaction and qubit bias. What happen for the
accessible parameters  of the present-day experiments? In addition,
VVP at small static bias $\varepsilon \le 1$  has not been discussed
either so far, which might however be more important.

Here, we calculate  energy levels in the VVP  for different static
bias $\varepsilon \le 1$, which  are exhibited in Fig. \ref{bias}.
Compare to the exact ones, one can find that VVP deviates strongly
with the increase of the tunneling parameters $\Delta $ in the a
wide coupling regime $g<0.5$, and become more pronounced at small
static bias. Especially, around the experimentally accessible
coupling strength around $g=0.1$, VVP becomes worse considerably.
Therefore a new analytical treatment is highly desirable.

\subsection{Analytical approximations at different levels}

In the framework of Eqs. (\ref{exact1}) and (\ref{exact2}), analytical
approximations can be performed systematically. First, as a zero-order
approximation (ZOA), we omit the off-diagonal terms and have
\begin{eqnarray*}
\left( m-g^2-\frac \varepsilon 2-E\right) c_m-\;D_{m,m}d_m &=&0, \\
-D_{m,m}c_m+\left( m-g^2+\frac \varepsilon 2-E\right) d_m &=&0.
\end{eqnarray*}
Nonzero coefficients will give the following equation
\[
\left( m-g^2-\frac \varepsilon 2-E\right) \left( m-g^2+\frac
\varepsilon 2-E\right) -D_{m,m}^2=0,
\]
The eigenvalues are then given by
\begin{equation}
E_{\pm }=m-g^2\mp \frac 12\sqrt{\varepsilon ^2+4D_{m,m}^2}
\label{EZero},
\end{equation}
The corresponding eigenstate is
\begin{equation}
\left| m\right\rangle _{\pm }\varpropto \left(
\begin{array}{l}
\left( -1\right) ^mD_{mm}\left| m\right\rangle _A \\
\left( m-g^2-\frac \varepsilon 2-E_{\pm }\right) \left| m\right\rangle _B
\end{array}
\right).
\end{equation}

\begin{figure}[tbp]
\includegraphics[width=12cm]{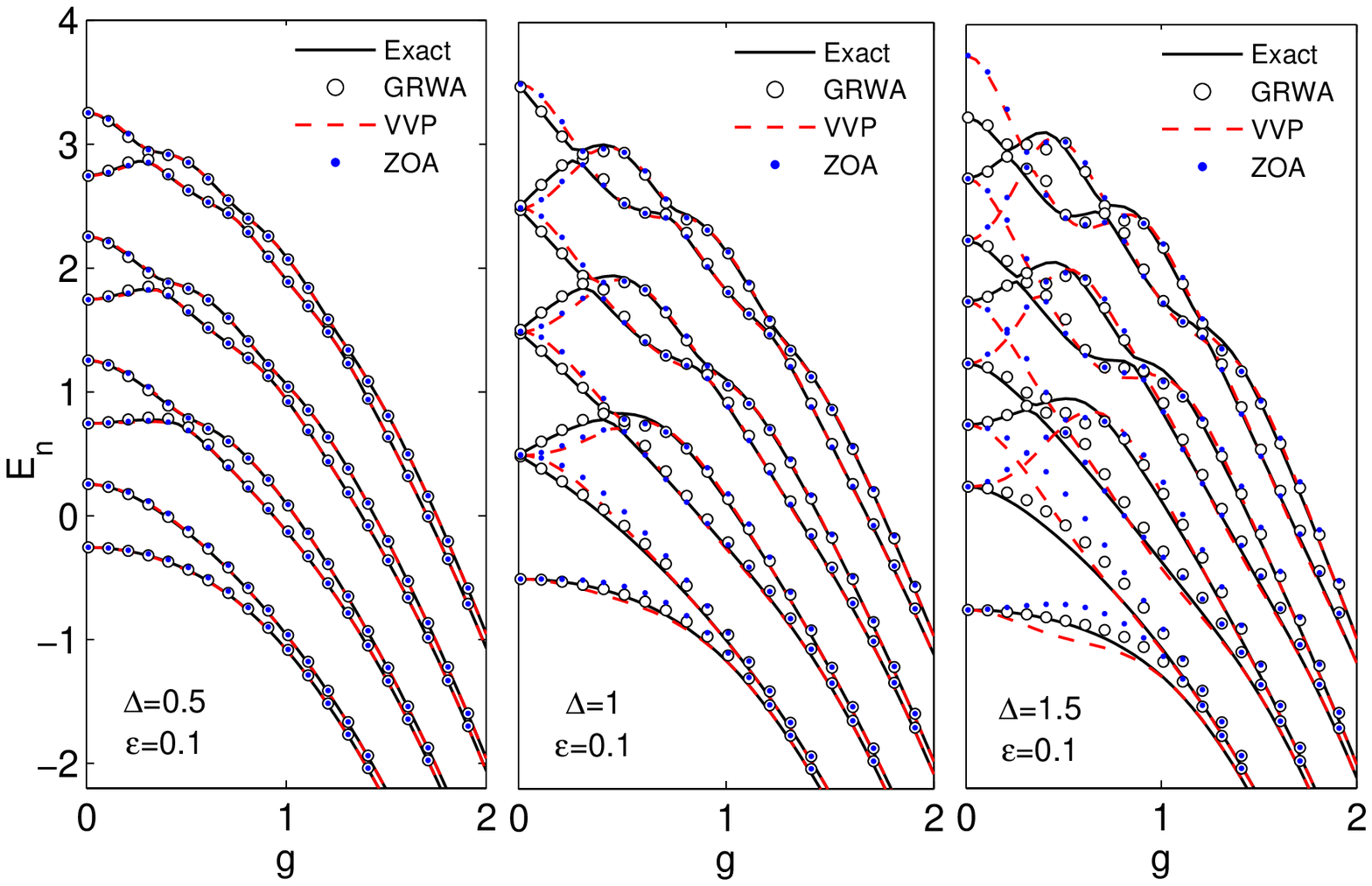}
\includegraphics[width=12cm]{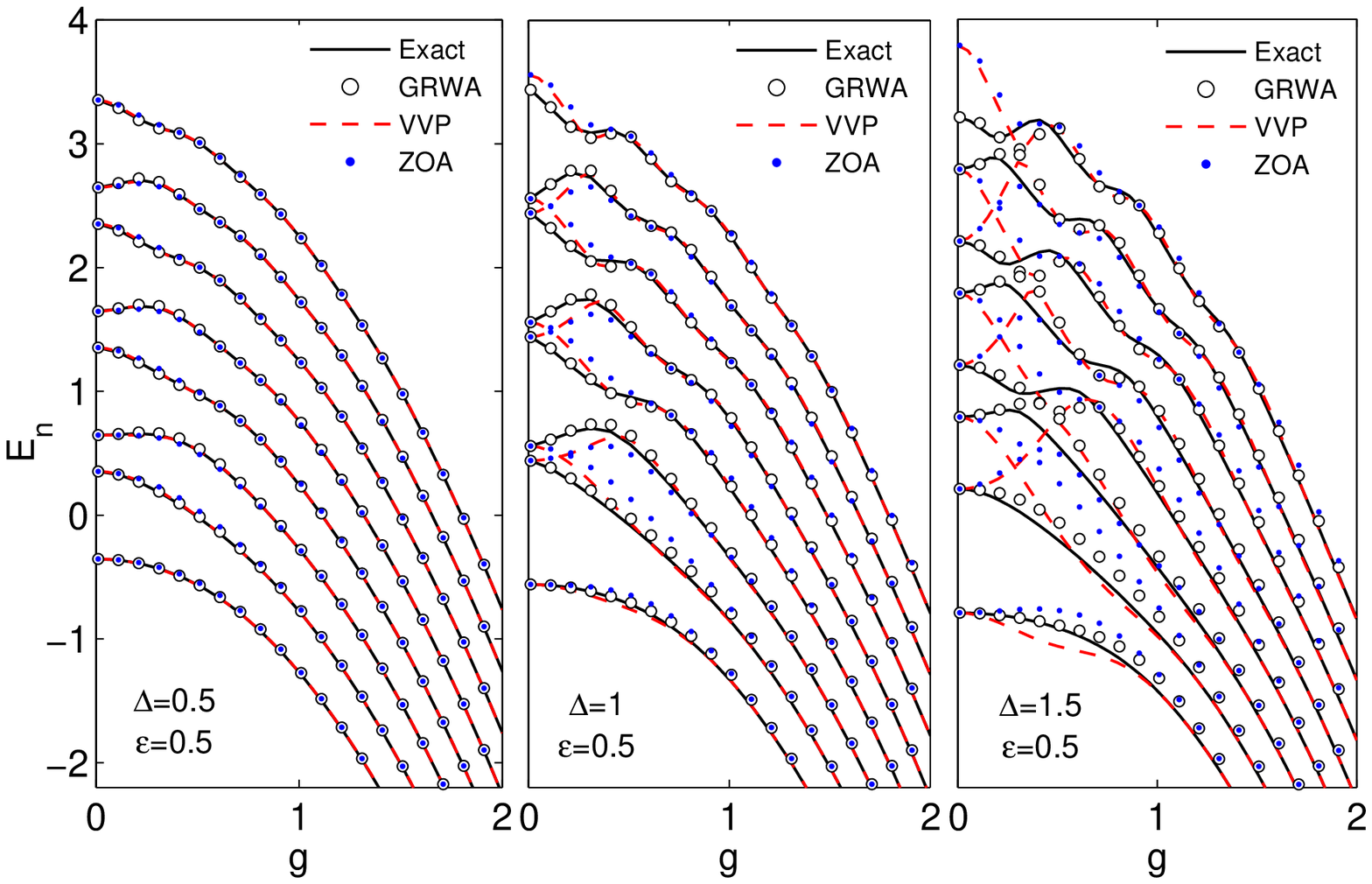}
\includegraphics[width=12cm]{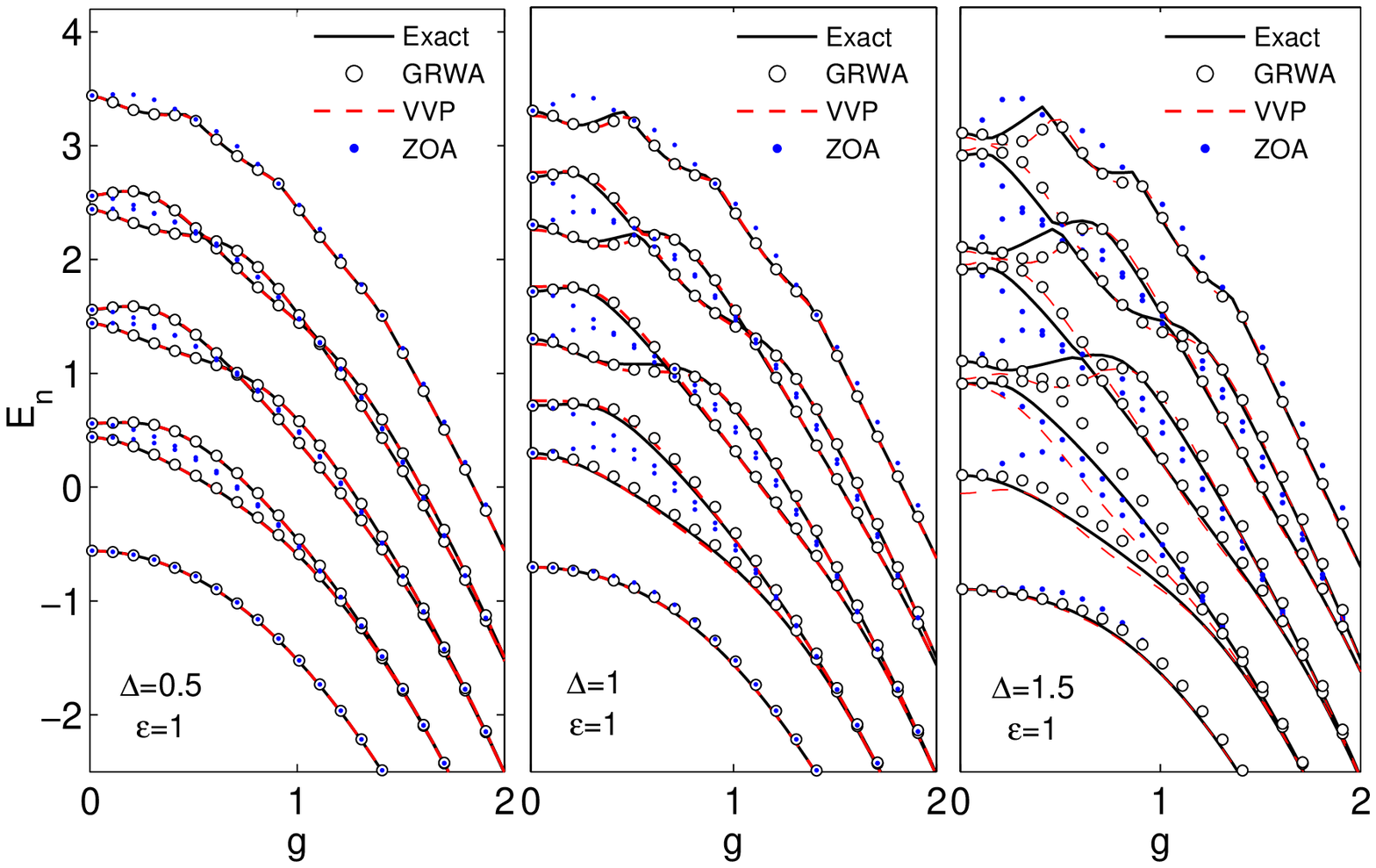}
\caption{ The energy levels as a function of coupling constant for
for different qubit bias $\epsilon=0.1$ (upper panel), $0.5$(middle
panel), and$ 1.0$ (down panel). The values of $\Delta$ are $0.5,
1.0$, and $1.5$ from left to right column. The present GRWA results
(black open circles) are compared to the exact (black solid lines),
VVP (red dashed lines), and ZOA (blue filled circles) ones. }
\label{bias}
\end{figure}

The ZOA energies with zero static bias are just the three terms
obtained in Eq. (\ref{DSC}). In Fig. \ref{bias}, we also plot the
ZOA energy levels against the coupling constant $g$ for several
static bias. It is demonstrated from the upper and middle panel that
for small static bias ( $ \varepsilon \leqslant 0.5$), ZOA is almost
equivalent to the VVP in all parameters. If the high accuracy is not
required, the simple expression of the eigensolutions in the ZOA
should be practically very useful, at least as a preliminary
estimate of some physical quantities .

The approximation can be easily improved step by step with the
consideration of more off-diagonal elements in the present
formalism. The first-order approximation (FOA) is performed by
selecting two coefficients $
\begin{array}{llll}
c_m & d_m & c_{m+1} & d_{m+1}
\end{array}
$. The determinants for any $m$ is given by
\begin{equation}
\left|
\begin{array}{llll}
\Omega _m^{-}(E) & -D_{mm} & 0 & -D_{m,m+1} \\
-D_{mm} & \Omega _m^{+}(E) & -D_{m,m+1} & 0 \\
0 & -D_{m+1,m} & \Omega _{m+1}^{-}(E) & -D_{m+1,m+1} \\
-D_{m+1,m} & 0 & -D_{m+1,m+1} & \Omega _{m+1}^{+}(E)
\end{array}
\right| =0  \label{Eq_first},
\end{equation}
where
\begin{equation}
\Omega _m^{\_}(E)=m-g^2-E-\frac \varepsilon 2,\;\Omega
_m^{+}(E)=m-g^2-E+\frac \varepsilon 2.
\end{equation}
Some roots of this quartic equation will give the energy levels. The
analytical expression might be a little bit complicate but should be given
unambiguously.

We first revisit the zero-bias case $\varepsilon =0$. In this case,
due to the parity symmetry, we can set $d_n=\pm c_n$, then both
equations give $ \left( m-g^2\right) c_m\mp \sum_{n=0\
}\;D_{mn}c_n=Ec_m$. In the FOA, the determinant takes the following
$2$-by-$2$ block form
\[
\left|
\begin{array}{ll}
\left( m-g^2-E\mp D_{m,m}\right) & \mp D_{m,m+1} \\
\mp D_{m+1,m} & \left( m+1-g^2-E\mp D_{m+1,m+1}\right)
\end{array}
\right| =0,
\]
where the sign $-(+)\;$is for even (odd) parity. We can readily have two
roots for even parity
\begin{equation}
E_m^{(1,2)}=m-g^2+\frac 12-\frac 12\left( D_{m+1,m+1}+D_{m,m}\right)
\pm \frac 12\sqrt{\left[ 1+\left( D_{m,m}-D_{m+1,m+1}\right) \right]
^2+4D_{m,m+1}^2}  \label{1m2roots},
\end{equation}
and other two roots for odd parity
\begin{equation}
E_m^{(3,4)}=m-g^2+\frac 12+\frac 12\left( D_{m+1,m+1}+D_{m,m}\right)
\pm \frac 12\sqrt{\left[ 1-\left( D_{m,m}-D_{m+1,m+1}\right) \right]
^2+4D_{m,m+1}^2} \label{2m2roots}.
\end{equation}

In the ansatz of the wavefunction (\ref{wavefunction}), the dimensions of
the Hilbert space is only $2(N_{tr}+1)$. So for each $m$, we only have two
eigenvalues for excited states. The other two roots for each $m$ should be
omitted. Note that at weak coupling, the parity for each eigenstate is fixed
and arranged from bottom to above with the order as the first even state,
then followed by two odd states, two even states, two odd states, and so on.
Therefore, the excited states $1$ and $2$ are of odd parity, which should be
given by $E_0^{(3,4)}$, the excited states $3$ and $4\;$ are of even parity,
then given by $E_1^{(1,2)}$, the excited states $5$ and $6\;$are of odd
parity, then given by $E_{2}^{(3,4)}$, $\;$and so on. In this way, two
eigenvalues for excited states for any $m$ can be summarized as
\begin{equation}
E_m=m-g^2+\frac 12+\frac{(-1)^m}2\left( D_{m+1,m+1}+D_{m,m}\right)
\pm \frac 12\sqrt{\left[ 1-(-1)^m\left( D_{m,m}-D_{m+1,m+1}\right)
\right] ^2+4D_{m,m+1}^2}  \label{2roots}.
\end{equation}
Besides, the GS energy is given by $E_0^{(1)}$
\begin{equation}
E_{GS}=\frac 12-g^2-\frac 12\left( D_{1,1}+D_{0,0}\right) -\frac
12\sqrt{ \left[ 1+\left( D_{0,0}-D_{1,1}\right) \right]
^2+4D_{0,1}^2} \label{GS_E}.
\end{equation}

The FOA results in Eqs. (\ref{2roots}) and (\ref{GS_E}) have been given
directly by the determinant with $2$-by-$2$ block form in Ref. \cite{liu} by
two present authors and one collaborator previously. We here display the
derivation in detail. Especially we rule out two pseud solutions for each $m$
by taking the fixed parity of the eigenstates into account.

Surprisingly Eq. (\ref{2roots}) is exactly the same as the previous
GRWA result Eq. (20) in Ref. \cite{Irish} by Irish.  We now become aware  that the
previous GRWA, which were obtained in an alternative way within a lengthy
derivation, is just FOA in the present scheme.  Actually this expression has been
derived much earlier within substantially different approaches\cite{Feranchuk}.
 What is more, we can straightforwardly perform the
second-order approximation for the further improvement, and
extension to the biased case in the present framework, which is
however not so easy to operate within Irish's approach. To the best
of our knowledge, GRWA with finite static bias does not exist in the
literature.

For the finite bias $\varepsilon \neq 0$, the parity symmetry is broken with
$\varepsilon ,\;$with the following notation
\begin{eqnarray}
E &=&x+m-g^2-\frac \varepsilon 2 \\
u &=&-D_{mm},v=-D_{m,m+1},w=-D_{m+1,m+1}  \nonumber,
\end{eqnarray}
The determinant can be reduced to
\begin{equation}
\left|
\begin{array}{llll}
-x & u & 0 & v \\
u & -x+\varepsilon & v & 0 \\
0 & v & 1-x & w \\
v & 0 & w & 1-x+\varepsilon
\end{array}
\right| =0.
\end{equation}
The corresponding quartic equation is
\[
x^4+bx^3+cx^2+dx+e=0,
\]
where
\begin{eqnarray*}
b &=&-2-2\varepsilon,  \\c &=&1+3\varepsilon +\varepsilon ^2-\left(
2v^2+u^2+w^2\right), \\ d &=&\left( 2v^2+u^2+w^2-\varepsilon
-1\right) \varepsilon +2\left( u^2+v^2\right),  \\e &=&\left(
uw-v^2\right) ^2-u^2\left( 1+\varepsilon \right) -v^2\varepsilon.
\end{eqnarray*}
The solutions to this quartic equation are given in the Appendix A.
Compared to the exact solutions, we find that the second and third
roots $x_2$ and $x_3$ in Eqs. (\ref{r2}) and (\ref{r3}) are
generally the true solutions. The GS energy is given by the first
root $x_1$ in Eq. (\ref{r1}) for $m=0$. We also call the FOA with
finite static bias as GRWA. In this way, we can calculate the
eigenenergies uniquely and straightforwardly, which are shown in
Fig. \ref{bias} with black circles. It is very interesting to find
that the present GRWA results are very close to the exact ones in
the whole coupling regime for wide range of the static bias. Compare
to the VVP at static bias $\varepsilon \le 1$, the present GRWA is
obviously much better.

As stated above, for zero-static bias case $\varepsilon =0$, there
is still room to improve by performing the higher order
approximation. In Ref. \cite{Irish}, after a unitary transformation,
only the "energy-conserving" one excitation terms like their Eq.
(18), a generalization of the energy-conserved term in the usual
RWA, is kept in their Eqs. (13) and (14), so it is called GRWA.
Because the present FOA is equivalent to GRWA, so in the
second-order approximation, the terms beyond their Eq. (18) must be
included, so we term this improvement to GRWA as beyond the RWA
(BRWA). In other words, BRWA can not be implemented within any
renormalized RWA form like in Ref. \cite{Irish}.

In the BRWA, the analytical expression can be uniquely and clearly
derived within the following procedure. The determinant is
\begin{equation}
\left|
\begin{array}{lll}
\left( m-g^2-E\mp D_{m,m}\right) & \mp D_{m,m+1} & \mp D_{m,m+2} \\
\mp D_{m,m+1} & \left( m+1-g^2-E\mp D_{m+1,m+1}\right) & \mp D_{m+1,m+2} \\
\mp D_{m,m+2} & \mp D_{m+1,m+2} & \left( m+2-g^2-E\mp D_{m+2,m+2}\right)
\end{array}
\right| =0,\label{BRWA}
\end{equation}
where $-(+)$ for even(odd) parity, which can be simplified as
\[
\left|
\begin{array}{lll}
-X\mp u & \mp x & \mp y \\
\mp x & \left( 1-X\mp v\right) & \mp z \\
\mp y & \mp z & \left( 2-X\mp w\right)
\end{array}
\right| =0,
\]
where
\begin{eqnarray*}
E &=&X+m-g^2, \\
u &=&D_{m,m},v=D_{m+1,m+1},w=D_{m+2,m+2}, \\
x &=&D_{m,m+1},y=D_{m,m+2},z=D_{m+1,m+2},
\end{eqnarray*}
which gives the following cubic equation
\begin{equation}
X^3+bX^2+cX+d=0  \label{eq_third},
\end{equation}
where
\begin{eqnarray*}
b &=&\left( u+v+w\right) -3, \\
c &=&-\left( x^2+y^2+z^2\right) +u\left( v-1\right) +\left( u+v-1\right)
\left( w-2\right), \\
d &=&\allowbreak \left( 2u-uw+y^2\right) \left( 1-v\right)
-z^2u+x^2\left( 2-w\right) +2xyz,
\end{eqnarray*}
for even parity, and
\begin{eqnarray*}
b &=&-\left( u+v+w\right) -3, \\
c &=&-\left( x^2+y^2+z^2\right) +u\left( 1+v\right) +\left( u+v+1\right)
\left( 2+w\right), \\
d &=&\allowbreak \left( y^2-2u-uw\right) \left( 1+v\right)
+z^2u+x^2\left( 2+w\right) -2xyz,
\end{eqnarray*}
for odd parity.

The three different roots to the cubic equation can be found in the
Appendix A. In this approximation, we have more than one eigenvalues
for each eigenstate with fixed parity, which are all true solutions
physically, but only some of them would be selected. The criterion
for the unique formulae for the BRWA is that the solutions are the
most close to the exact results in the whole parameter regime. In
this way, we find that, for even (odd) number $m$, two roots $y_1$ and
$y_2$ in Eqs. (\ref{y1}) and (\ref{y2}) of the determinant with odd
(even) parity would generally give the best eigenvalues for the
excited states.  The GS state is given by the first root $y_1$ of
the  $m=0$ determinant with even parity.  Actually, Eq. (\ref{BRWA})
has been  written out in Ref. \cite{liu} by
two present authors and one collaborator. But the detailed expression
for the eigenvalues was not presented. Even the further third approximation
was also performed in Ref. \cite{liu}. The direct comparisons
between these different order approximations to the GRWA\cite{Irish,Feranchuk} 
have not been given, which however could reveal the advantage of this scheme.

We examine the BRWA energy levels against the qubit-oscillator
detuning $ \delta$ for fixed couplings $g=0.1,0.5,1.0$, and $1.5$,
respectively in Fig. \ref{unbias}, where the GRWA results have been also collected.
It is interesting to note that 
BRWA result  is always more close to the
exact one than GRWA  one in all values of the coupling strength, which
becomes more pronounced with increasing $\delta $.

Due to the counter-rotating wave terms, the eigenfunctions and
eigenvalues of the JC  model without the RWA present an open problem
because they are not known in anything like a closed form, even
given the exact solutions reported recently\cite{QingHu1,Braak}. No
analytical expressions for the exact eigenvalues are available in
the literature, to the best of our knowledge. The analytical
expressions presented in this paper, which is not exact but work
well, might be practically useful.

\section{summary}

In this paper, by an effective scheme within two displaced bosonic
operators with equal positive and negative displacements, we study
the qubit-oscillator systems analytically in an unified way. Many
previous analytical treatments, such as GRWA, an expansion in the
qubit tunneling matrix element in the deep strong coupling regime
can be recovered in the present scheme. More over, we extend the
GRWA to the finite-bias case. The results is much better than VVP in
the weak and intermediate coupling regime, which is more
experimentally interesting. For the zero static bias, the GRWA is
further improved to BRWA, which is more close to the exact ones at
large detuning while the GRWA deviates strongly. The analytical
expression is explicitly given for future applications.

\section{ACKNOWLEDGEMENTS}

This work was supported by National Natural Science Foundation of China,
National Basic Research Program of China (Grant Nos. 2011CBA00103 and
2009CB929104). \appendix

\section{Solutions to univariate cubic and quartic equations}

The univariate cubic equation can be generally expressed as
\begin{equation}
x^3+bx^2+cx+d=0.  \nonumber
\end{equation}
Its solutions can be found in any Mathematics manual. If
\[
\Gamma =B^2-4AC<0,
\]
with
\[
A=b^2-3c,B=bc-9d,C=c^2-3bd,
\]
there are three different real roots
\begin{eqnarray}
y_1 &=&\frac{-b-2\sqrt{A}\cos \theta }3, \\ \label{y1} y_2
&=&\frac{-b-2\sqrt{A}\cos \left( \theta -\frac{2\pi }3\right) }3, \\
\label{y2} y_3 &=&\frac{-b-2\sqrt{A}\cos \left( \theta +\frac{2\pi
}3\right) }3, \label{y3}
\end{eqnarray}
where
\begin{equation}
\theta =\frac 13\arccos \left( \frac{2Ab-3B}{2\sqrt{A^3}}\right).
\end{equation}

The univariate quartic equation can be generally expressed as
\begin{equation}
x^4+bx^3+cx^2+dx+e=0,  \nonumber
\end{equation}
Its four solutions are exactly the four solutions of the following two
quadratic equations
\begin{eqnarray}
x^2+\frac{b+z}2x+\left( y+\frac{by-d}z\right) &=&0, \\
x^2+\frac{b-z}2x+\left( y-\frac{by-d}z\right) &=&0,
\end{eqnarray}
where $z=\sqrt{8y+b^2-4c}$ and $y$ is the third root $y_3$ in Eq. (\ref{y3})
of the following cubic equation
\[
y^3-\frac c2y^2+\left( \frac{bd}4-e\right) y+\frac{e\left(
4c-b^2\right) -d^2 }8=0.
\]
We haver checked that $\Gamma <0$ in all parameters in the present case.
Therefore the four roots are
\begin{eqnarray}  \label{r3}
x_1 &=&-\frac 14\left( b+z\right) -\frac 14\sqrt{\left( b+z\right)
^2-\frac{ 16y\left( b+z\right) -16d}z}, \\  \label{r1} x_2 &=&-\frac
14\left( b+z\right) +\frac 14\sqrt{\left( b+z\right) ^2-\frac{
16y\left( b+z\right) -16d}z}, \\  \label{r2}  x_3 &=&-\frac 14\left(
b-z\right) -\frac 14\sqrt{\left( b-z\right) ^2+\frac{ 16y\left(
b-z\right) -16d}z}, \\  \label{r3}  x_4 &=&-\frac 14\left(
b-z\right) +\frac 14\sqrt{\left( b-z\right) ^2+\frac{ 16y\left(
b-z\right) -16d}z}.
\end{eqnarray}

\end{document}